\documentclass[aps,prl,groupedaddress,fleqn,twocolumn]{revtex4}
\pdfoutput=1
\usepackage{graphicx}
\usepackage{amsmath}
\usepackage{gensymb}
\begin{document}

\title{Emergence and evolution of $k$-gap in spectra of liquid and supercritical states}
\author{C. Yang$^{1}$}
\author{M. T. Dove$^{1}$}
\author{V. V. Brazhkin$^{2}$}
\author{K. Trachenko$^{1}$}
\address{$^1$ School of Physics and Astronomy, Queen Mary University of London, Mile End Road, London, E1 4NS, UK}
\address{$^2$ Institute for High Pressure Physics, RAS, 108840, Troitsk, Moscow, Russia}

\begin{abstract}
Fundamental understanding of strongly-interacting systems necessarily involves collective modes, but their nature and evolution is not generally understood in dynamically disordered and strongly-interacting systems such as liquids and supercritical fluids. We report the results of extensive molecular dynamics simulations and provide direct evidence that liquids develop a gap in solid-like transverse spectrum in the reciprocal space, with no propagating modes between zero and a threshold value. In addition to the liquid state, this result importantly applies to the supercritical state of matter. We show that the emerging gap increases with the inverse of liquid relaxation time and discuss how the gap affects properties of liquid and supercritical states.
\end{abstract}


\maketitle

Dynamical and thermodynamic properties of an interacting system are governed by collective excitations, or modes. Collective modes have been studied in depth and are well-understood in solids and gases. This
is not the case for the liquid state where the combination of strong interactions and dynamical disorder has been thought to preclude the development of a general theory \cite{landau} including understanding
the nature of collective modes.

Collective modes in solids include one longitudinal and two transverse acoustic modes. In gases, the collective mode is one longitudinal long-wavelength sound wave considered in the hydrodynamic approximation. In liquids, collective modes are well-understood in the hydrodynamic approximation $\omega\tau<1$ \cite{hydro}, where $\omega$ is frequency and $\tau$ is liquid relaxation time, the average time it takes for a molecule to diffuse the distance equal to interatomic separation \cite{frenkel,fntau}. Importantly, there is a different regime of wave propagation: $\omega\tau>1$ where waves propagate in the constant-structure environment, i.e. in the solid-like regime. Experiments have reported both indirect and direct evidence for the existence of solid-like waves in liquids and have ascertained that they are essentially different from the hydrodynamic modes \cite{mon-na,rec-review,ruocco,grim,scarponi,burkel,water-fast,pilgrim,water-tran,hoso,mon-ga,sn,hoso3} including those discussed in generalized hydrodynamics \cite{boon,baluca}.

The first proposal regarding solid-like waves in liquids was due to Frenkel \cite{frenkel} who proposed that at times smaller than $\tau$, particles do not jump and hence the system behaves like a solid. Therefore, for frequencies larger than

\begin{equation}
\omega_{\rm F}=\frac{1}{\tau}
\label{tau}
\end{equation}

\noindent the liquid supports two transverse acoustic modes as does the solid (glass or crystal). The longitudinal acoustic mode is unmodified (except for different dissipation laws in regimes $\omega\tau>1$ and $\omega\tau<1$ \cite{frenkel}): density fluctuations exist in any interacting medium, and in liquids they have been shown to propagate with wavelengths extending to the shortest interatomic separation \cite{mon-na,burkel,rec-review,mon-ga,sn,hoso3}.

The proposal that liquids are able to support solid-like transverse modes with frequencies extending to the highest frequency implies that liquids are similar to solids in terms of collective
excitations. Therefore, main liquid properties such as energy and heat capacity can be described using the same first-principles approach based on collective modes as in solids, an assertion that was
considered as unusual in the past when no evidence for propagating solid-like modes in liquids existed. Importantly, high-frequency modes are particularly relevant for liquid thermodynamics because, similarly
to solids, they make the largest contribution to system's energy and other properties whereas the contribution of hydrodynamic modes is negligible \cite{ropp}.

Observed in viscous liquids (see, e.g., Refs. \cite{grim,scarponi}), high-frequency transverse modes were later studied in low-viscosity liquids on the basis of positive dispersion \cite{burkel,rec-review,water-fast,water-tran}. More recently, high-frequency transverse modes were directly measured in the form of distinct dispersion branches and verified in computer modeling \cite{hoso,mon-na,mon-ga,water-tran,sn,hoso3}. This has been done at constant temperature and $\tau$.

Although Eq. (\ref{tau}) has been the traditional basis for understanding solid-like transverse modes in liquids \cite{dyre}, the crucial question is whether the frequency gap actually emerges in the liquid transverse spectrum as Eq. (1) and common wisdom \cite{dyre} predict and as more recent work \cite{newpaper} seems to suggest? Does this gap change with temperature (pressure) as $\frac{1}{\tau}$? Answering these questions directly is essential for fundamental understanding of collective modes in liquids and for liquid theory in general.

A recent detailed analysis \cite{ropp} predicts the following dispersion relationship for liquid transverse modes:

\begin{equation}
\omega=\sqrt{c^2k^2-\frac{1}{\tau^2}}
\label{k}
\end{equation}

\noindent where $k$ is the absolute value of the vector in the reciprocal space (wave vector), $c$ is the speed of transverse sound and $\tau$ is understood to be the full period of particles' jump motion equal to twice Frenkel's $\tau$.

Interestingly and differently from (\ref{tau}), Eq. (\ref{k}) predicts that liquid transverse acoustic modes develop a {\it gap in the reciprocal space}, between $0$ and $k_{\rm gap}$:

\begin{equation}
k_{\rm gap}=\frac{1}{c\tau}
\label{k1}
\end{equation}

Eqs. (\ref{k})-(\ref{k1}) further predict that the $k$-gap increases with temperature because $\tau$ decreases.

It is interesting to discuss why the gap develops in $k$-space rather than in the frequency domain. Eq. (\ref{k}) follows from our solution of the Navier-Stokes equation extended by Frenkel to include the solid-like elastic response of liquids at time shorter than $\tau$. This gives a wave equation with dissipation, from which (\ref{k}) follows \cite{ropp}. A qualitatively similar result can be also inferred from generalized hydrodynamics where the hydrodynamic transverse current correlation function is generalized to include large $k$ and $\omega$ \cite{boon}. The approach assumes that the shear viscosity function $K$, the memory function for transverse current correlation function, exponentially decays with time $\tau$, giving a resonant frequency in the correlation function. If we now identify $K$ at short times with $c^2$, the resonant frequency becomes similar to (\ref{k}). A gap in $k$-space, albeit different from (\ref{k1}), is also mentioned in a different method \cite{bryk}. Using (\ref{k1}), we write the condition $k>k_{\rm gap}$ approximately as $\lambda<d_{\rm el}$, where $d_{\rm el}=c\tau$ is liquid elasticity length, the propagation length of a shear wave in the liquid \cite{elast}. The microscopic meaning of $d_{\rm el}$ follows from noting that liquid particles jump with a period of $\tau$ and hence disrupt the wave continuity at distances larger than $c\tau$, setting the longest wavelength of propagating waves. Therefore, the condition $k>k_{\rm gap}$ in Eq. (\ref{k}) is consistent with the condition that allowed wavelengths should be smaller than the wave propagation length \cite{fn}.

Importantly, we predict that the $k$-gap also emerges in the supercritical state of matter, the state which has traditionally been viewed as a gray area on the phase diagram with unknown properties intermediate between gases and liquids. We have earlier proposed that solid-like transverse modes should propagate in supercritical fluids below the Frenkel line (FL) \cite{ropp,prl,phystoday,pre}. We therefore predict that supercritical fluids below the FL should also develop the same gap (\ref{k1}) in the transverse spectrum.

The main aim of this paper is to obtain direct evidence for the gap discussed above. We perform extensive molecular dynamics simulations in different types of liquids and supercritical fluids, including noble and molecular. We find that a gap develops in solid-like transverse acoustic spectrum in reciprocal space which increases with the inverse of liquid relaxation time. These specific results call for new high-temperature and pressure experiments.

We aimed to study the propagation of solid-like transverse waves in liquids with different structure and bonding types and have performed molecular dynamics simulations of noble liquid Ar and molecular CO$_2$ \cite{supp1}.
The pressure was fixed at 40 bar for subcritical Ar, 10 kbar for supercritical Ar and 9 kbar for supercritical CO$_2$. The temperature was extended well above critical for the last two systems.

We calculate the propagating transverse modes from transverse current correlation functions \cite{baluca}: $C({\bf k},t)=\frac{k^2}{N}\langle J_x({-\bf k},t)J_x({\bf k},0)\rangle=\frac{k^2}{N}\langle
J_y(-{\bf k},t)J_y({\bf k},0)\rangle$, where $N$ is number of the particles, transverse currents $J({\bf k},t)=\sum_{j=1}^{N}{\bf k}\times {\bf v}_j(t)${exp}$[-i{\bf k}\cdot {\bf r}_j(t)]$, ${\bf v}$ is
particle velocity and wavevector ${\bf k}$ is along the z-axis. The spectra of transverse currents are calculated as the Fourier transform of the real part of $C({\bf k},t)$ (the imaginary part of $C({\bf k},t)$ is calculated to be zero within the error as expected). A smoothing function is often used for the analysis of $C(k,t)$ in order to reduce the noise \cite{sn,hoso3}. To get better quantitative and model-free results, we choose not to use the smoothing. Instead, we repeat our simulations 20 times using different starting velocities and average the results. This produces $C(k,t)$ with reduced noise which does not change when the number of simulations is increased. We show examples of $C(k,\omega)$ for different peak frequencies in the Supplemental Material.

Our main observation is related to the evolution of dispersion curves. We plot intensity maps $C(k,\omega)$ in Figure \ref{3} and observe that a {\it gap} develops in $k$ space and the range of transverse modes progressively shrinks. A maximum of $C(k,\omega)$ at frequency $\omega$ is related to a propagating mode at that frequency and gives a point ($k,\omega$) on the dispersion curve \cite{baluca}. We plot dispersion curves in Figure {\ref{2} and observe a detailed evolution of the gap. At the highest temperature simulated, $C(k,\omega)$ becomes not easily discernable from the noise.

\begin{figure*}
\begin{center}
{\scalebox{0.78}{\includegraphics{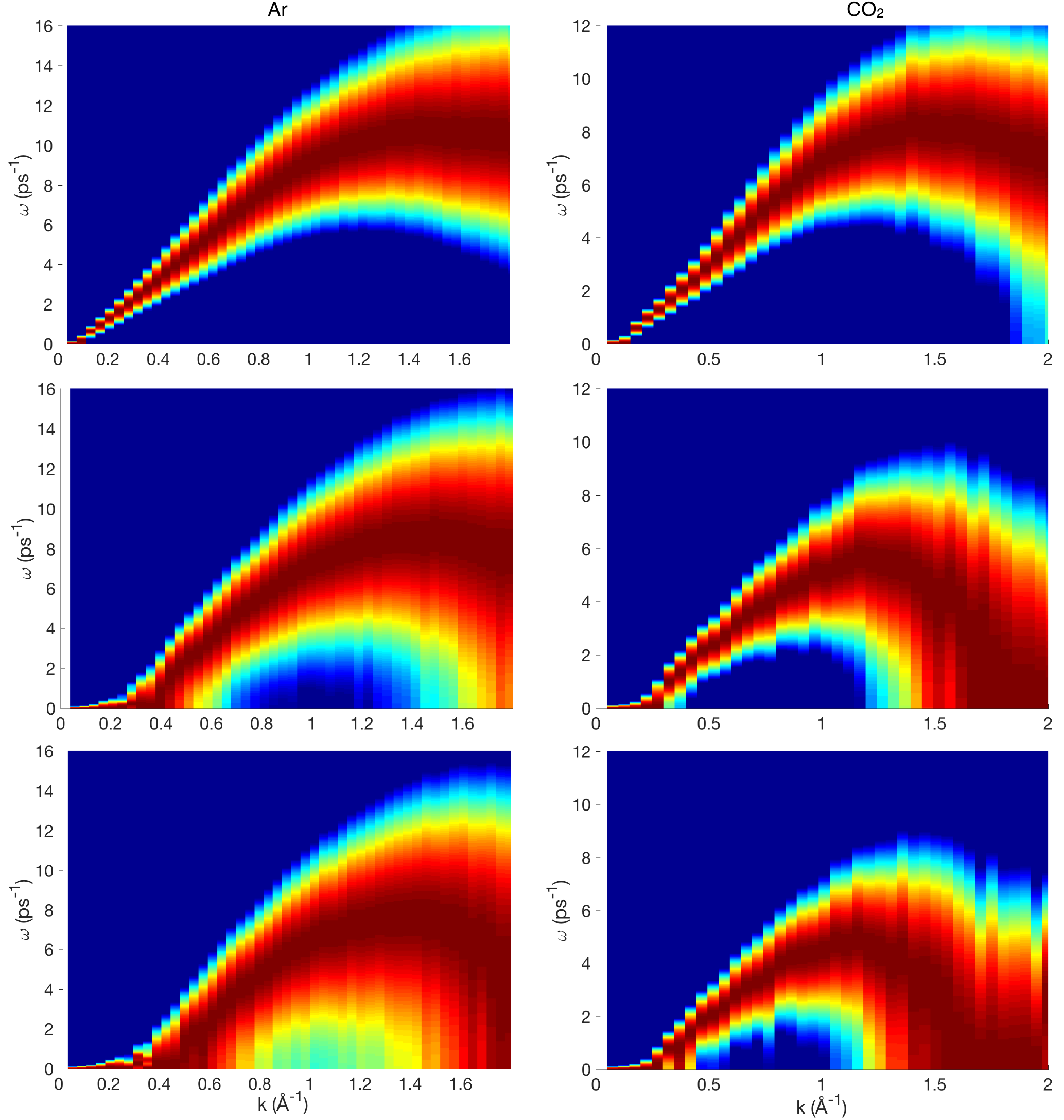}}}
\end{center}
\caption{Intensity maps of $C(k,\omega)$ for supercritical Ar at 250 K (top), 350 K (middle) and 450 K (bottom) and supercritical CO$_2$ at 300 K (top), 400 K (middle) and 500 K (bottom).
The maximal intensity corresponds to the middle points of dark red areas and reduces away from them.}
\label{3}
\end{figure*}

\begin{figure}
\begin{center}
{\scalebox{0.635}{\includegraphics{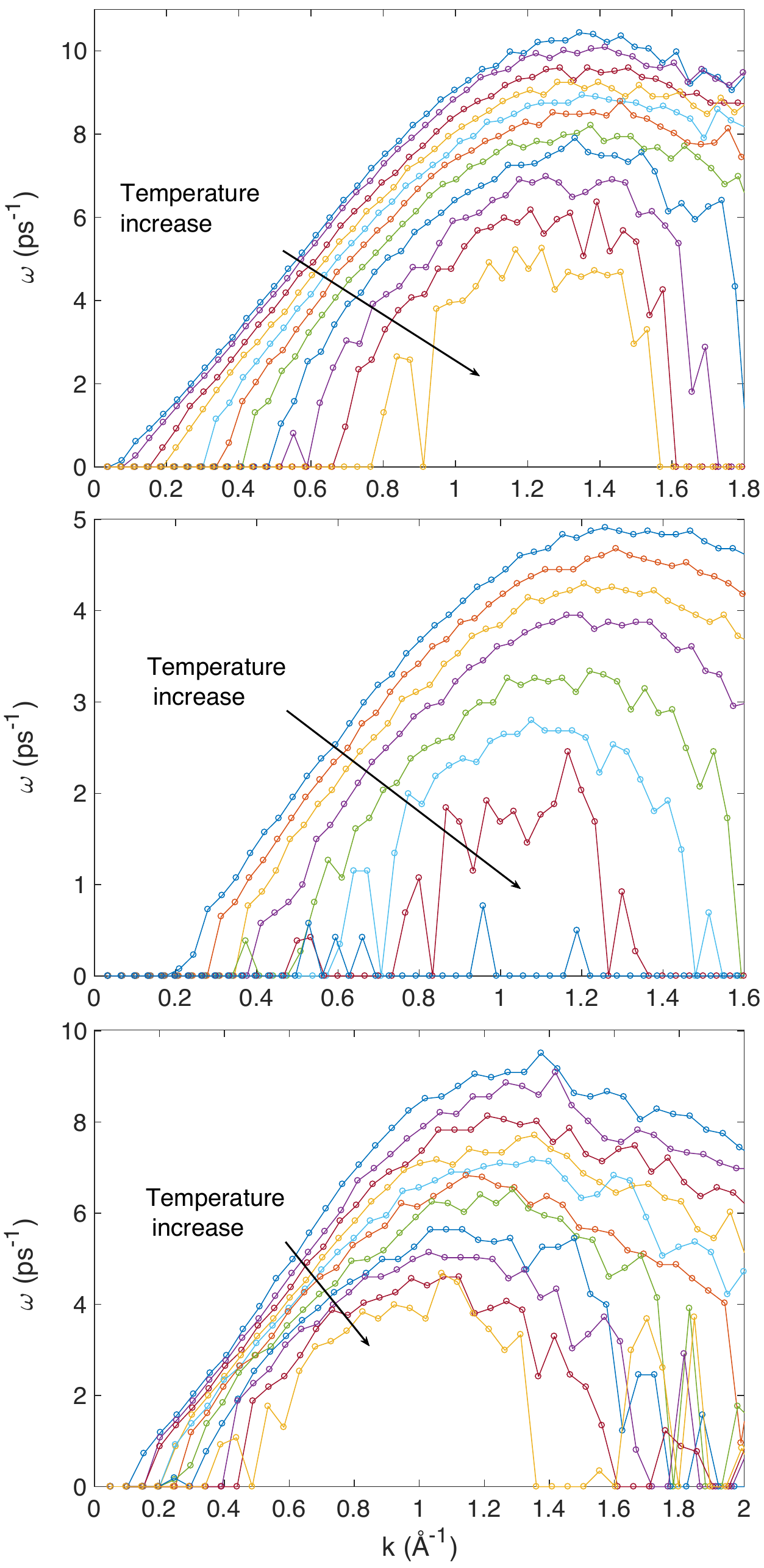}}}
\end{center}
\caption{Phonon dispersion curves of supercritical Ar at 200-500 K and 550 K (top), subcritical liquid Ar at 85-120 K (middle) and supercritical CO$_2$ at 300-600 K (bottom). The temperature increment is 30 K, 5 K and 30 K for supercritical Ar, subcritical liquid Ar and supercritical CO$_2$, respectively. The deviation from linearity (curving over) of dispersion curves at large $k$ is related to probing the effects comparable to interatomic separations (in the solid this correspond to curving over of $\omega\propto\sin ck$ at large $k$). This effect is not accounted for in the theory leading to the gap in (\ref{k}) and (\ref{k1}) because the theory is formulated in the continuous medium \cite{ropp} and therefore describes the $k$-gap in the linear part of the dispersion law. We show $k$ in the range slightly extending the first pseudo-zone boundary (FPZB) at low temperature. As volume and interatomic separation increase with temperature, FPZB shrinks, resulting in the decrease of $\omega\propto\sin ck$ at large $k$ beyond the FPZB. This effect is unrelated to the $k$-gap.}
\label{2}
\end{figure}

We observe that the gap $k_{\rm gap}$ develops in all systems simulated. Importantly, the simulated systems where we detect transverse modes extend into the supercritical state: our maximal temperature and pressure correspond to (205.6$P_c$, 6.3$T_c$) for Ar and (122.0$P_c$, 2.0$T_c$) for CO$_2$. It has remained unclear whether the supercritical state is able to support solid-like transverse modes at all but we have recently proposed that the supercritical state supports transverse modes below the Frenkel line (FL), the line that demarcates liquid-like and gas-like properties on the phase diagram \cite{prl,ropp,phystoday,pre}. Below the Frenkel line, particle motion consists of both oscillatory and diffusive components. Above the line, the oscillatory component of particle motion is lost, leaving only diffusive motion as in a gas. Approaching the line from below approximately corresponds to $\tau$ becoming equal to the shortest period of transverse modes, at which point the system becomes depleted of all available transverse modes according to (\ref{tau}). Using the previously calculated FL for Ar \cite{prl} and CO$_2$ \cite{fr-co2}, we find that propagating solid-like transverse modes reported in Figure 2 correspond to supercritical Ar and CO$_2$ below the FL. 

We note that the intensity of $C(k,\omega)$ peaks decreases with temperature for all mode frequencies, but lower-frequency peaks decay much faster as compared to higher-frequency ones. In examples shown in the first figure in the Supplemental Material, both 2 THz and 8 THz transverse modes show a clear peak at 200 K but whereas the peak of the 2 THz mode almost disappears at 350 K, the 8 THz mode peak remains pronounced. This is consistent with the experiments showing that low-frequency transverse phonons are not detected \cite{mon-na,mon-ga}.

We also note that reduced peak intensity of $C(k,\omega)$ at very high temperature, together with the persisting noise, can obfuscate the criterion of a propagating mode because a difference between a peak in $C(k,\omega)$ at low temperature and a broad shoulder at high temperature becomes less pronounced. However, one can also consider the oscillatory behavior of $C(k,t)$ as an indicator of a propagating mode. Shown in second figure in the Supplemental Material, $C(k,t)$ for $k$ close to the Brillouin pseudo-boundary has minima and oscillatory behavior at 900 and 950 K but not at 1000 K, even though $C(k,\omega)$ shows no maxima in the temperature range 900-1000 K. In agreement with this, the temperature of the Frenkel line demarcating propagating and non-propagating transverse modes is about 1000 K \cite{prl}.

We can now directly verify the predictions for the gap $k_{\rm gap}=\frac{1}{c\tau}$ in (\ref{k1}). First, in Fig. {\ref{4} we observe either a nearly linear relationship or a correct trend between $k_{\rm gap}$ and $\frac{1}{\tau}$ (more computationally consuming CO$_2$ with smaller cell size involves smaller resolution of $k$ and larger noise). The increase of slope of $k_{\rm gap}$ vs $\frac{1}{\tau}$ at large $\frac{1}{\tau}$ at high temperature is expected because $c$ decreases with temperature ($\frac{1}{c}$ increases). Second, we calculate $c$ for each system from the dispersion curves in the linear regime at small $k$ in Figure \ref{2} and find them to be in reasonable agreement with $c$ extracted from the linear regime in Figure \ref{4} for the three systems studied.

\begin{figure}
\begin{center}
{\scalebox{0.5}{\includegraphics{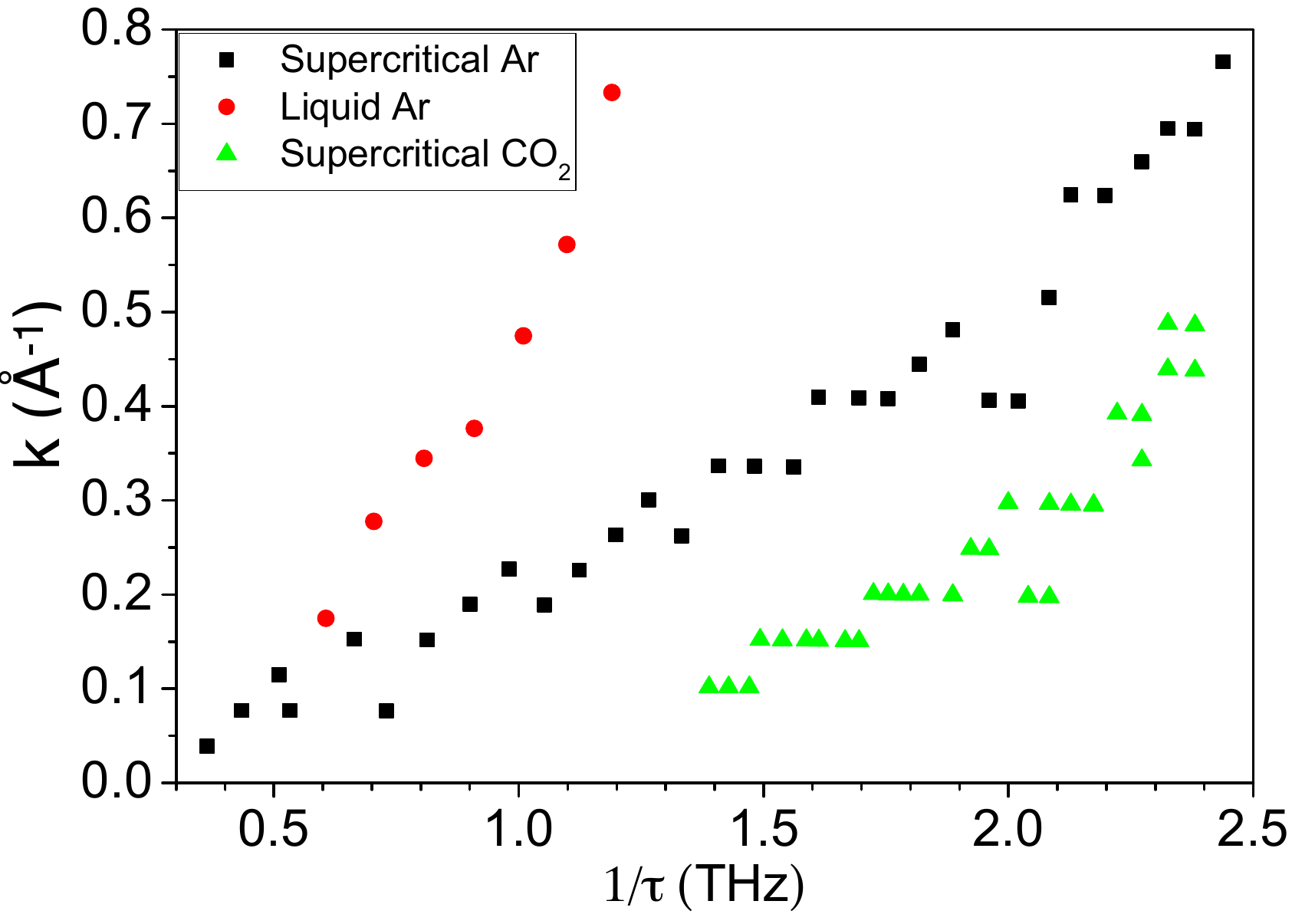}}}
\end{center}
\caption{The width of $k$-gap vs $\frac{1}{\tau}$ for subcritical liquid Ar in the range 85-115 K, supercritical Ar in the range 200-500 K and supercritical CO$_2$ in the range 300-600 K. 
}
\label{4}
\end{figure}

Our results are important for understanding liquid thermodynamics. The $k$-gap implies that the energy of transverse modes can be calculated as $E_t=\int\limits_{k_{\rm gap}}^{k_{\rm D}}E(k)\frac{6N}{k_{\rm D}^3}k^2dk$, where $N$ is the number of particles, $k_{\rm D}$ is Debye wavenumber and the factor $\frac{6N}{k_{\rm D}^3}$ is due to $2N$ transverse modes between 0 and $k_{\rm D}$ in the solid. Taking $E(k)=k_{\rm B}T$ in the classical case gives

\begin{equation}
E_t=2Nk_{\rm B}T\left(1-\left(\frac{\omega_{\rm F}}{\omega_{\rm D}}\right)^3\right)
\label{ene2}
\end{equation}

\noindent where $\omega_{\rm D}=ck_{\rm D}$ is Debye frequency.

The same result can be obtained in the Debye model if we calculate the energy of transverse modes propagating above the frequency $\omega_{\rm F}$ as Eq. (\ref{tau}) predicts, i.e. if we consider a gap in the {\it frequency} spectrum. Indeed, this energy can be written as $\int\limits_{\omega_{\rm F}}^{\omega_{\rm D}}g(\omega)k_{\rm B}Td\omega$, where $g(\omega)=\frac{6N}{\omega_{\rm D}^3}\omega^2$ is Debye density of states of transverse modes. This gives the same $E_t$ as in (\ref{ene2}). As $\omega_{\rm F}$ increases with temperature, the number of transverse modes decrease, resulting in the decrease of specific heat in agreement with the experimental results for many liquids and supercritical fluids in a wide temperature range \cite{ropp,ling}. Hence from the point of view of thermodynamics, the transverse modes can be considered to have a frequency gap $\omega_{\rm F}$, in agreement with the original assumption (\ref{tau}).

Writing (\ref{k}) as

\begin{equation}
E=\sqrt{p^2c^2-E^2_{\rm F}}
\label{last}
\end{equation}

\noindent where $E=\hbar\omega$ and $E_{\rm F}=\hbar\omega_{\rm F}$, we see that quasiparticles with energy $E_{\rm F}$ act as filters to suppress the quasiparticle excitations whose energy $pc$ is below $E_{\rm F}$. We propose that the energy-momentum relationship (\ref{last}) may be of interest in other areas of physics including quantum field theory.

In summary, we showed that collective modes in liquids and supercritical fluids develop a $k$-gap in the solid-like transverse spectrum.

We are grateful to the Royal Society, CSC and V. V. B. to RSF 14-22-00093.

\section{Supplemental Material}

Interatomic potentials were optimized to reproduce liquid properties in a wide range of pressure and temperature. We have used the DL$\_$POLY molecular dynamics program \cite{dlpoly} and systems with 125,000 and 117,912 particles for Ar and CO$_2$, respectively, with periodic boundary conditions. The empirical potential for Ar was the common Lennard-Jones potential with parameters $\epsilon=0.01032$ eV and $\sigma=3.4$ \AA. The empirical potential for CO$_2$ included intermolecular Buckingham interactions $U(r)=A\exp(-r/\rho)-B/r^6$, rigid molecular units and electric charges for CO$_2$ with electric charges \cite{co2} with parameters included in the table.

\begin{table}[h]
\caption{Parameters and charges for CO$_2$ potential.}
\begin{tabular}{p{2cm} p{2.5cm} p{2.5cm} p{2.5cm} p{2.5cm}}
\hline
&A (eV)&$\rho$ (\AA)&C (eV$\cdot$ \AA$^6$)&Charge ($e$) \\
\hline
C--O & 1978.66 & 0.2637 & 12.61 & \\
O--O & 2109.85 & 0.2659 & 22.28 & -0.30403\\
C--C & 1122.96 & 0.2778 & 0.0 & 0.60806\\
\hline
\end{tabular}
\end{table}

We equilibrated the system during 15 ps in NPT ensemble and observed that the fluctuations of system properties (e.g., volume and temperature) do not change with time during the subsequent 10 ps run. We collected the results during the following 200 ps for Ar, and 80 ps for CO$_2$ in NVE ensemble. The averaging of $C(k,t)$ involved 20 simulations with different sets of initial velocities randomly generated and conforming to a Gaussian distribution each. $\tau$ was calculated as the decay time of the intermediate scattering function by a factor of $e$ \cite{decay}.\\

\begin{figure}[h!]
\begin{center}
{\scalebox{0.47}{\includegraphics{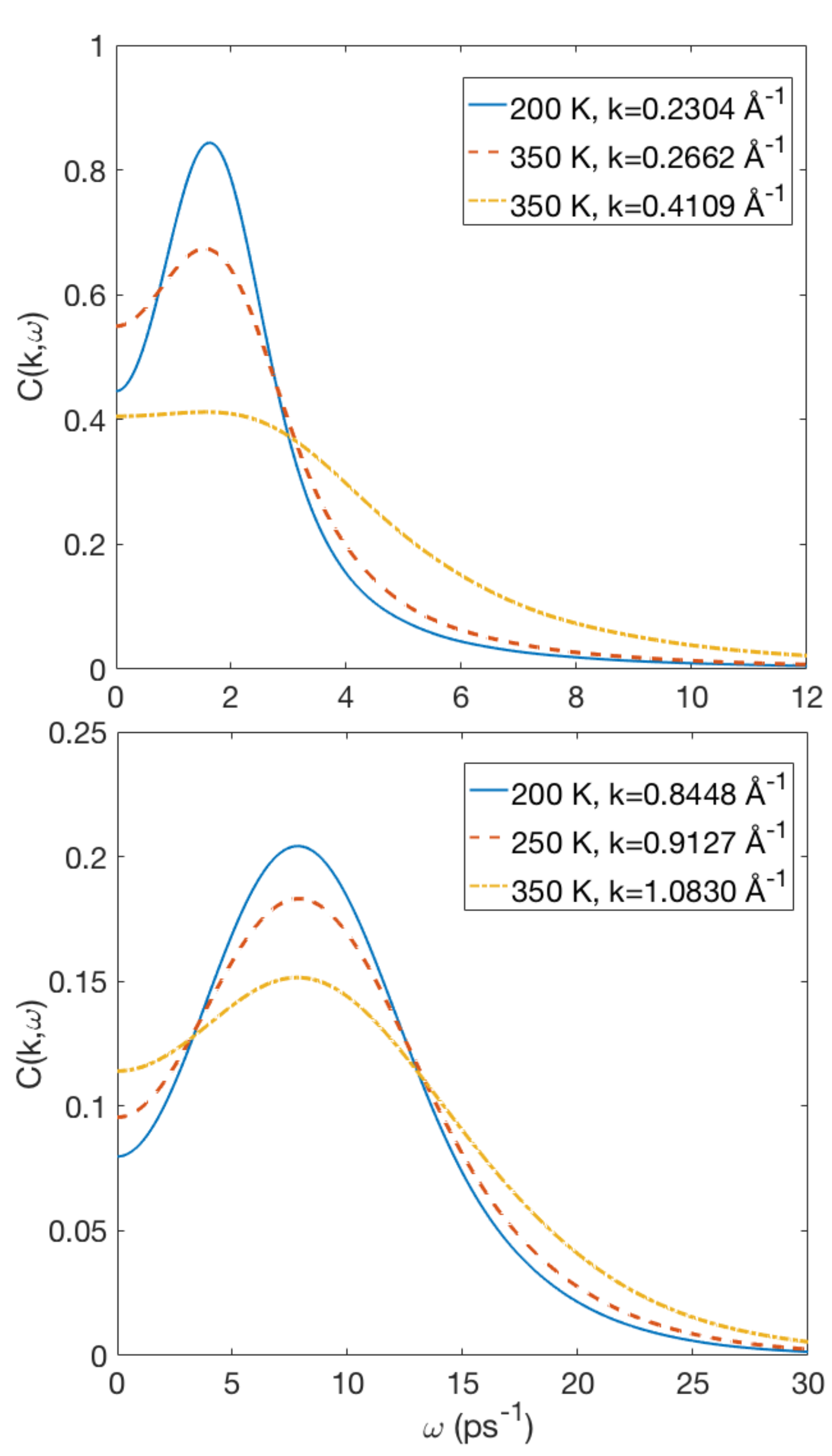}}}
\end{center}
\caption{$C(k,\omega)$ of supercritical Ar at 200 K, 250 K and 350 K for two phonon modes with peak frequencies of about 2 THz (top) and 8 THz (bottom). $C(k,\omega)$ at higher temperature are shown at larger $k$ because $\omega$ decreases with temperature (see Fig. 2 in main text).}
\label{1}
\end{figure}

\begin{figure}[h!]
\begin{center}
{\scalebox{0.3}{\includegraphics{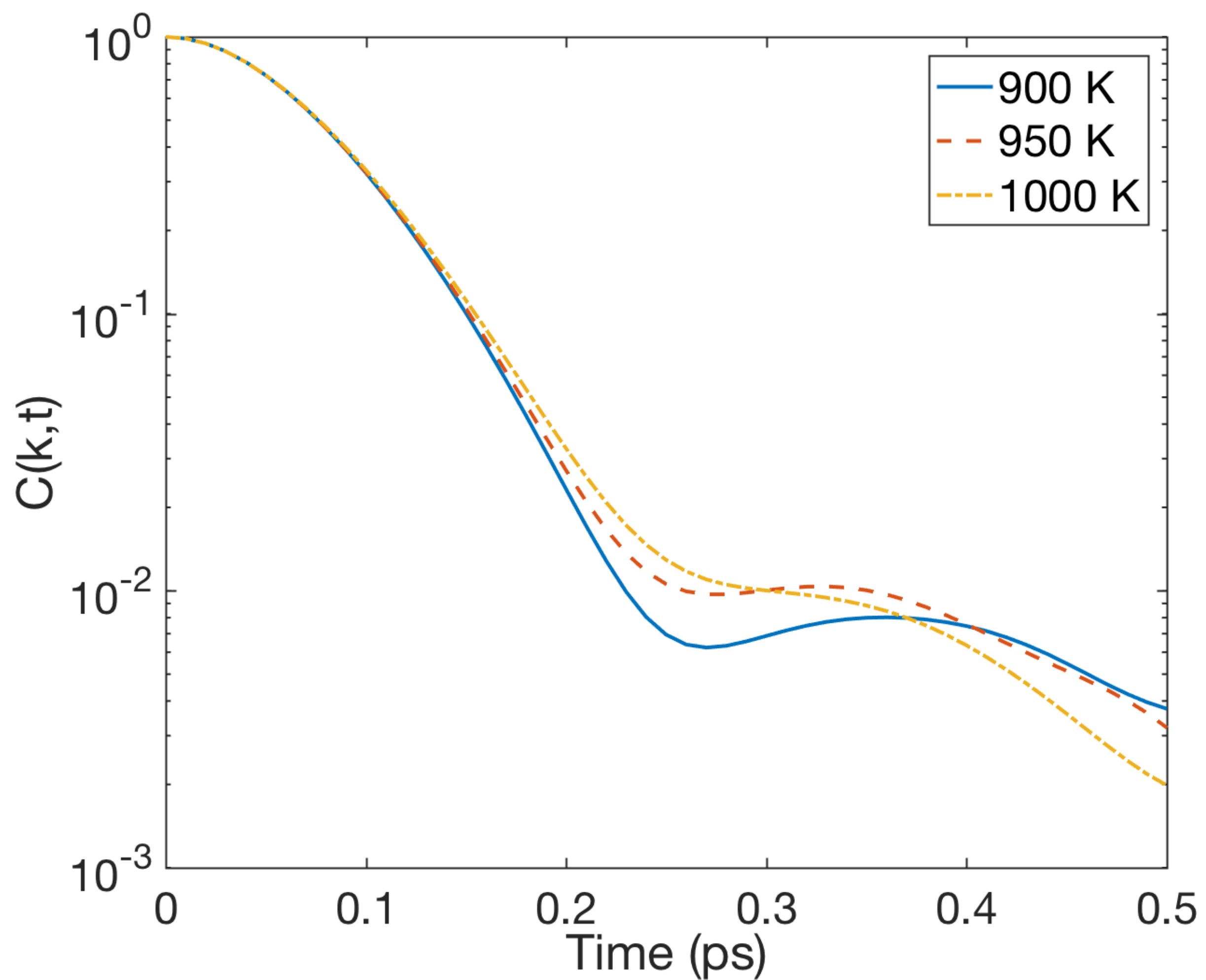}}}
\end{center}
\caption{$C(k,t)$ of supercritical Ar near the first Brillouin pseudo-zone boundary showing the crossover from the oscillatory to monotonic behavior.}
\label{31}
\end{figure}


\begin{thebibliography}{99}
\bibitem{fntau} $\tau$ is readily measured in experiments including dielectric relaxation experiments, NMR, positron annihilation spectroscopy and can be derived from viscosity data.
\end{thebibliography}
\end{document}